\documentclass[a4paper,showpacs,amssymb,pre,superscriptaddress]
              {revtex4}

\usepackage{graphicx}

\begin{document}

\title{Current and efficiency enhancement in Brownian motors
driven by non Gaussian noises.}

\author{Sebastian Bouzat}\email{bouzat@cab.cnea.gov.ar}
\affiliation{Grupo de F\'{\i}sica Estad\'{\i}stica,\\
Centro At\'omico Bariloche (CNEA) and Instituto Balseiro
(UNC-CNEA)\\
8400 San Carlos de Bariloche, R\'{\i}o Negro, Argentina}
\author{H. S. Wio$^{1,}$}\email{wio@ifca.unican.es}
\affiliation{Instituto de F\'{\i}sica de Cantabria,
         Avda. Los Castros s/n \\
         E-39005 Santander, Spain}

\begin{abstract}
We study Brownian motors driven by colored non Gaussian noises,
both in the overdamped regime and in the case with inertia, and
analyze how the departure of the noise distribution from Gaussian
behavior can affect its behavior. We analyze the problem from two
alternative points of view: one oriented mainly to possible
technological applications and the other more inspired in natural
systems. In both cases we find an enhancement of current and
efficiency due to the non-Gaussian character of the noise. We also
discuss the possibility of observing an enhancement of the mass
separation capability of the system when non-Gaussian noises are
considered.
\end{abstract}


\maketitle

\section{Introduction}

The study of noise induced transport by "ratchets" has attracted
in recent years the attention of an increasing number of
researchers due to the biological interest and also to its
potential technological applications \cite{inic,inic2}. Since the
pioneering works, besides the built-in ratchet-like bias and
correlated fluctuations (see for instance \cite{magnasco}),
different aspects have been studied, such as tilting
\cite{tilting,tilting2} and pulsating \cite{pulsating} potentials,
velocity inversions \cite{tilting,invers}, etc. There are some
relevant reviews \cite{review1,review2} where the biological
and/or technological motivation for the study of ratchets can be
found.

Recent studies on the role of non Gaussian noises on some
noise-induced phenomena like stochastic resonance, resonant
trapping, and noise-induced transitions
\cite{qRE1,qRE1p,qRE2,qruido1,qruido2,qruido3} have shown the
possibility of strong effects on the system's response. For
instance, enhancement of the signal-to-noise ratio in
stochastic resonance, enhancement of the trapping current in
resonant trapping, or shifts in the transition line for
noise-induced transitions. These results motivate the interest in
analyzing the effect of non Gaussian noises on the behavior of
Brownian motors. Here we analyze the effect of a particular class
of colored non Gaussian noise on the transport properties of
Brownian motors. Such a noise source is based on the nonextensive
statistics \cite{tsallis,tsallis2} with a probability distribution
that depends on $q$, a parameter indicating the departure from
Gaussian behavior: for $q = 1$ we have a Gaussian distribution,
and different non Gaussian distributions for $q > 1$ or $q < 1$.

Some of the motivations for studying the effect of non Gaussian
noises are, in addition to its intrinsic interest within the realm
of noise induced phenomena, the existence of experimental data
indicating that for several biological problems fluctuations have
a non Gaussian character. Examples are current measurements
through voltage-sensitive ion channels in a cell membrane or
experiments on the sensory system of rat skin \cite{nature}. Also,
recent detailed studies on the source of fluctuations in different
biological systems \cite{caltech} clearly show that, in such a
context, noise sources in general are non Gaussian. Even though
the previous arguments refer to biological aspects that are not
directly related to ratchets, they strongly induce to think
about the possible relevance of considering non Gaussian noises in
those biological situations where the ratchet transport mechanism
can play a role. In addition, from the point of view of
technological applications, the finding of new conditions that may
lead to an enhancement of the efficiency of the devices is always
desirable.

We show here that, as a consequence of the non Gaussian character
of the driving noise and from two alternative points of view, we
can find a kind of enhancement of the system's response. The
first --direct-- point of view, takes as free parameters those that
could be controlled in the case of technological applications. In
this case we find a remarkable increase of the current together
with an enhancement of the motor efficiency when non-Gaussian
noises are considered (showing an optimum for a given degree of
departure from the $q=1$ Gaussian behavior). Moreover, when
inertia is taken into account it is found that, again when
departing from the Gaussian case, there seems to be a remarkable
increment in the mass separation capability of these devices. The
second point of view is the more natural one when thinking of
biological systems, as it considers the non-Gaussian noise as a
primary source. In this case we also find an enhancement of the
current and efficiency due to departure from Gaussian behavior,
which occurs for low values of noise intensity.

We begin presenting the general framework within which we will
work, and the nature of the non Gaussian noise. We continue
discussing the first of the two points of view, and the results
showing the enhancement we can find within it. After that we
discuss the second point of view where we compare Gaussian and non
Gaussian behaviors but adopting a constant width criterion, and
discuss the results. Finally we draw some general conclusions.

\section{Framework}

We begin considering the general system
\begin{equation}
\label{xpto} m\frac{d^2 x}{dt^2}=-\gamma\frac{dx}{dt}
-V'(x)-F+\xi(t)+\eta(t),
\end{equation}
where $m$ is the mass of the particle, $\gamma$ the friction
constant, $V(x)$ the ratchet potential, $F$ is a constant ``load''
force, and $\xi(t)$ the thermal noise satisfying $\langle \xi(t)
\xi(t') \rangle= 2 \gamma T \delta(t-t')$. Finally, $\eta(t)$ is
the time correlated forcing (with zero mean) that allows the
rectification of the motion, keeping the system out of thermal
equilibrium even for $F=0$. For this type of ratchet model several
different kind of time correlated forcing have been considered in
the literature \cite{review1,review2}. In almost all studies
authors have used Gaussian noises. The few exceptions which
considered non Gaussian processes correspond mainly to the case of
dichotomic noises \cite{inic2,tilting,review2}.

The main characteristic introduced by the non Gaussian form of the
forcing we consider here, is the appearance of arbitrary strong
``kicks'' with relatively high probability when compared, for
example, with the Gaussian Ornstein--Uhlenbeck (OU) noise and, of
course, with the dichotomic non Gaussian process.

We will consider the dynamics of $\eta(t)$ as described by the
following Langevin equation \cite{qRE1}
\begin{equation}
\label{etapto} \frac{d\eta}{dt}=-\frac{1}{\tau}\frac{d}{d \eta}
V_q(\eta) + \frac{1}{\tau} \zeta(t),
\end{equation}
with $\langle\zeta(t)\rangle=0$ and
$\langle\zeta(t)\zeta(t')\rangle=2D\delta(t-t')$, and
$$V_q(\eta) = \frac{D}{\tau(q-1)}\,
\ln[1+\frac{\tau}{D}(q-1)\frac{\eta^2}{2}].$$ Previous studies of
such processes in connection with stochastic resonance problems
\cite{qRE1,qRE1p} and dynamical trapping \cite{qruido1}, have
shown that the non Gaussian behavior of the noise leads to
remarkable effects. For $q=1$, the process $\eta$ coincides with
the OU one (with a correlation time equal to $\tau$), while for
$q\neq 1$ it is a non Gaussian process. As shown in \cite{qRE1},
for $q<1$ the stationary probability distribution has a bounded
support, with a cut-off at $|\eta|= \omega \equiv
[(1-q)\tau/(2D)]^{-\frac{1}{2}}$, with a form given by
\begin{equation}
P_q(\eta)=\frac{1}{Z_q}
\left[1-(\frac{\eta}{\omega})^2\right]^\frac{1}{1-q},
\end{equation}
for $|\eta|<\omega$ and zero for $|\eta|>\omega$ ($Z_q$ is a
normalization constant). Within the range $1<q<3$, the probability
distribution is given by
\begin{equation}
P_q(\eta)=\frac{1}{Z_q} \left[1+\frac{\tau (q-1)\eta^2}{2
D}\right]^\frac{1}{1-q}
\end{equation}
for $-\infty<\eta<\infty$, and decays as a power law (slower than
a Gaussian distribution). Finally, for $q>3$, this distribution
can not be normalized.

Hence, we see that keeping $D$ constant, the width or dispersion
of the distribution increases with $q$. This means that, the
higher the $q$, the stronger the ``kicks'' that the particle will
receive. Figure 1 depicts the typical form of this distribution
for $q$ smaller, equal and larger than 1.

\begin{figure}
\centering
\includegraphics[width=8cm]{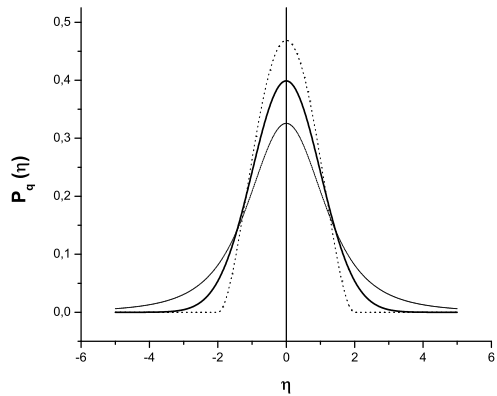}
\label{FigWio-1-2}
\caption{PDF vs $\eta$ for different values of $q$, and for
$D/\tau = 1.$ From top to bottom, the dotted line is for $q=0.5$,
the thick continuous line is for $q=1$ (Gaussian case), and the
thin continuous line is for $q=1.5$.}
\end{figure}

In \cite{qRE1} it was also shown that the second moment of the
distribution (which can be interpreted as the ``intensity" of the
non Gaussian noise) is given by
\begin{equation}\label{eqn5}
D_{ng} \equiv \langle \eta^2 \rangle =\frac{2D}{\tau (5-3q)},
\end{equation}
and diverges for $q \ge 5/3$. For the correlation time $\tau_{ng}$
of the process $\eta(t)$, we were not able to find an analytical
expression. However, it is known \cite{qRE1} that for $q \to 5/3$
it also diverges as $\sim (5-3q)^{-1}$. In our analysis, we will
only consider values of $q<5/3$, in order to keep finite values of
$D_{ng}$ and $\tau_{ng}$.

To conclude this Section, we briefly sketch some results about the
correlation time $\tau_{ng}$. As commented above, both the second
moment of $P_q(\eta)$ and the correlation time of the process
$\eta$, diverge for $q \to \frac{5}{3}$. For the second moment,
$D_{ng}$, we have the exact result shown in Eq. (\ref{eqn5}),
while for the correlation time, $\tau_{ng}$, we have no analytical
expression. However, in \cite{qRE1} we have observed, numerically,
the behavior of the correlation function
\begin{equation}
C(t) = \frac{\langle \eta (t+t') \eta (t')\rangle}{\langle \eta
(t') \eta(t')\rangle},
\end{equation}
in the stationary regime $t'\to \infty$. We have found that the
exponential decay of the correlations in the OU process is still
valid for $q < 1$ where we can write $C_q(t) \simeq
\exp(-t/\tau_{ng})$. This exponential behavior fails for $q > 1$
where, on the other side, $C_q(t)$ can be approximated by a
``q-exponential" \cite{tsallis} as
\begin{equation}
C_q(t) \simeq \left [ 1+(q-1)\frac{t}{\tau_{ng}} \right ]
^{\frac{1}{1-q}}.
\end{equation}
The characteristic correlation time $\tau_{ng}$ defined as
\begin{equation}
\tau_{ng} = \int_0^{\infty} dt C_q(t),
\end{equation}
was shown to diverge for $q \to 5/3$ as $\tau_{ng} \approx
2/(5-3q)$.

In Fig. 2 we show, after simulation results for the correlation
function $C_q$, the dependence of $\tau_{ng}$ on $q$, as well as
the comparison with different approximations as follows
\begin{itemize}
\item  $\tau_{ng} \simeq \frac{2 \, \tau}{(5-3q)},$ \item
$\tau_{ng} \simeq \frac{2 \, \tau \, (2-q)}{(5-3q)},$ \item
$\tau_{ng} \simeq \frac{[1+ 4 (q-1)^{2}] \, \tau}{(5-3q)},$
\end{itemize}
where the third, ad hoc, approximation corresponds to the best fit
to the numerical data within the range $0.8 < q <5/3$. This
fitting result will be the one we will exploit after discussing
the most direct approach.

\begin{figure}
\centering
\includegraphics[width=8cm]{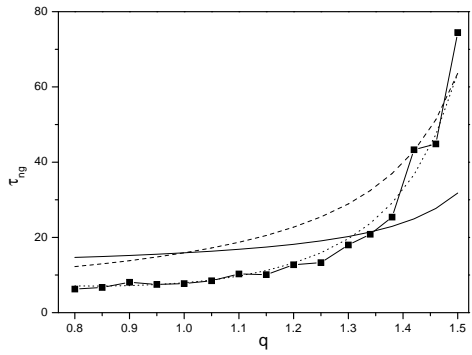}
\label{fig1}
\caption{Comparison of simulation and approximate
results for $\tau_{ng}$ vs $q$. Here $D=1$ and $\tau = 15.9$. Here
$\blacksquare$ correspond to the numerical data (the line joining
the points is only to guide the eye), the broken line is for the
first approximation, the continuous line for the second, and the
dotted line for the third fit.}
\end{figure}

\section{Case I: Constant $D$ and $\tau$.}

As our first approach we will analyze the results for the current
and efficiency as function of $q$ for constant values of $D$ and
$\tau$. These are the parameters that could be adequately
controlled, for example, in a designed technological device. The
analysis of the current and efficiency behavior for constant $D_{ng}$
and $\tau_{ng}$, that will be discussed in the next Section,
could be considered an even more relevant item --as $D_{ng}$
and $\tau_{ng}$ are parameters of the non Gaussian noise when we
think such noise as the ``primary source'' acting on the Brownian
particle.

In \cite{tilting}, essentially the same system described by Eq.
(\ref{xpto}) was analyzed for the case in which the time
correlated forcing belongs to a general family of stochastic
processes (non Gaussian in general) called ``kangaroo processes''.
Also in that work, and for the same kind of processes, a parameter
called the ``flatness'' of the noise was defined as the ratio of
the fourth moment to the square of the second moment of the
stationary distribution. The dependence of the current on the
flatness for the limit of small correlation time was discussed. It
is interesting to note that, as defined in \cite{tilting}, the
flatness of the stationary distribution of the process indicated
by Eq. (\ref{etapto}) is infinite for $q>7/5=1.4$, since the
fourth moment of the distribution diverges. As we will see, this
corresponds to the parameter region where we observe a decay in
the efficiency of the Brownian motor. However, it has to be noted
that the process here considered is not a ``kangaroo process''.

\subsection{Overdamped System}

Firstly, we analyze the overdamped regime setting $m=0$ and
$\gamma=1$. For the ratchet potential in this case we will
consider the same form as in \cite{magnasco} (with period $L=2 \pi$)
\begin{equation}
\label{V1dex} V(x)=V_1(x)=- \int^x dx' \left( \frac{\exp[\alpha
\cos(x')]}{J_0(i\alpha)} - 1 \right),
\end{equation}
where $J_0(i\alpha)$ is the Bessel function, and $\alpha=16$. The
form of $V_1(x)$ is shown in Fig. 3.a. The integrand in
Eq.(\ref{V1dex}) is the ratchet force ($-V'$) appearing in
Eq.(\ref{xpto}).

\begin{figure}
\centering
\includegraphics[width=12cm]{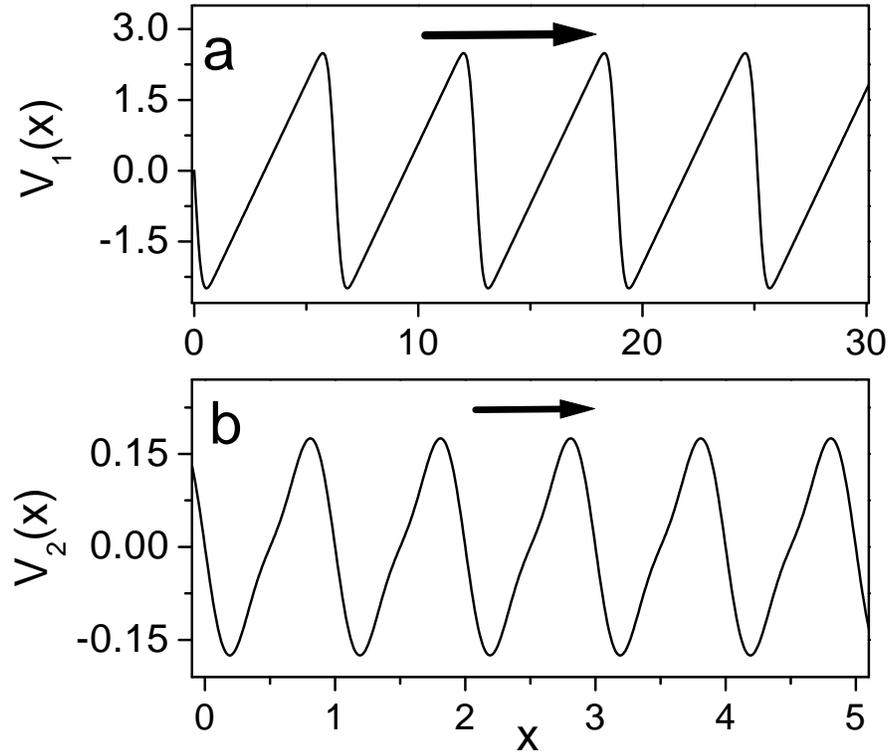}
\caption{The ratchet potentials considered in this work: $V_1(x)$
(a) and $V_2(x)$ (b). The arrows indicate the direction of the
current in normal situations (that is, when there is no current
``inversion").}
\end{figure}

We are interested on analyzing the dependence of the mean current
$J=\langle \frac{dx}{dt} \rangle/L$ and the efficiency
$\varepsilon$  on the different parameters, in particular, their
dependence on the parameter $q$. The efficiency is defined as the
ratio of the work (per unit time) done by the particle ``against''
the load force $F$
$$\lim_{T_f \to \infty}
\frac{1}{T_f}\int_{x=x(0)}^{x=x(T_f)} F dx(t),$$ to the mean power
injected to the system through the external forcing $\eta$
$$\lim_{T_f \to \infty} \frac{1}{T_f}\int_{x=x(0)}^{x=x(T_f)}
\eta(t) dx(t).$$ For the numerator we get $F\langle \frac{dx}{dt}
\rangle = F J L$, while for the denominator we obtain
$$\lim_{T_f
\to \infty} \frac{1}{T_f}\int_0^{T_f} \eta (t) \frac{dx}{dt} dt =
\frac{1}{\gamma}(\langle -V' \eta \rangle+\langle \eta^2 \rangle
).$$
Simulations show that the time average of $V'(x(t))\eta(t)$
is negligible in the latter equality (it is always several orders
of magnitude lower than $\langle \eta^2 \rangle$) and we may
approximate the denominator as $\langle \eta^2 \rangle/\gamma = 2
D [\gamma \tau (5-3q)]^{-1}$. Interesting and complete discussions
on the thermodynamics and energetics of ratchet systems can be
found in \cite{Seki}.

In the overdamped regime we are able to give an approximate
analytical solution for the problem, which is expected to be valid
in the large correlation time regime ($\frac{\tau}{D} \gg 1$): we
perform the adiabatic approximation of solving the Fokker-Planck
equation associated to Eq. (\ref{xpto}) assuming a constant value
of $\eta$ \cite{comFP}, analogous to the one used in \cite{new}.
This leads us to obtain an $\eta$--dependent value of the current
$J(\eta )$ that is then averaged over $\eta$ using the
distribution $P_q(\eta)$ \cite{qRE1} with the desired value of $q$
$$J = \int d\eta \, J(\eta ) \, P_q(\eta).$$

In Fig. 4, we show typical analytical results for the current and
the efficiency as functions of $q$ together with results coming
from numerical simulations (for the complete system given by
Eqs.(\ref{xpto}) and (\ref{etapto})). Calculations have been done
in a parameter region similar to that studied in
\cite{magnasco} but considering (apart from the difference
provided by the non Gaussian noise) a non--zero load force that
leads to a non--vanishing efficiency. As can be seen, although
there is not a quantitative agreement between theory and
simulations, the adiabatic approximation predicts qualitatively
very well the behavior of $J$ and $\varepsilon$ as $q$ is varied.
As shown in the figure, the current grows monotonously with $q$
(at least for $q<5/3$) while there is an optimal value of $q$ ($ >
1$) which gives the maximum efficiency. This fact could be
interpreted as follows: when $q$ is increased, the width of the
$P_q(\eta)$ distribution grows and high values of the non Gaussian
noise become more frequent, leading to an improvement of the
current. Although the mean value of $J$ increases monotonously
with $q$, the efficiency decays when $q$ approaches $5/3$ since
the denominator in the definition of $\varepsilon$, which is
essentially the dispersion of the noise distribution, diverges .

\begin{figure}
\centering
\includegraphics[width=10cm]{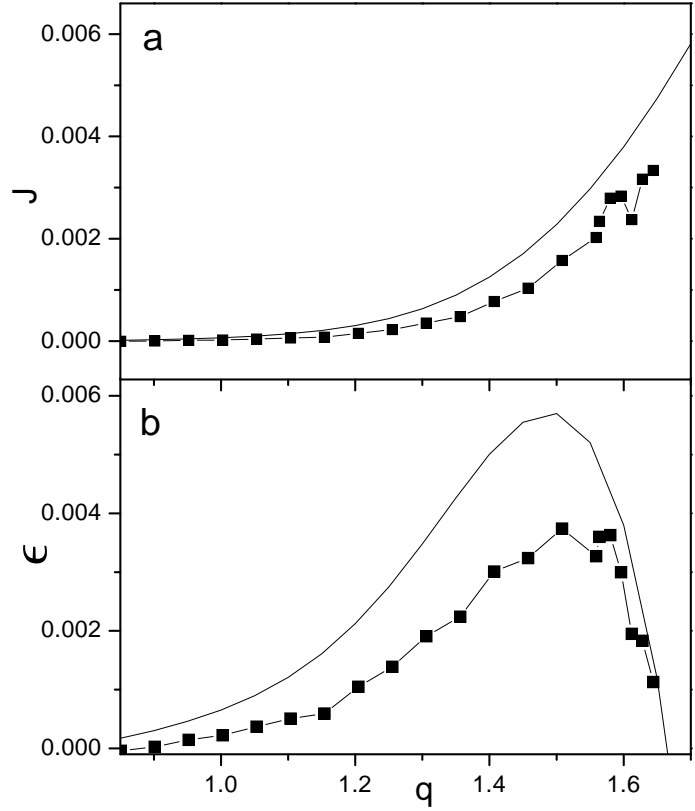}
\caption{Current (a) and efficiency (b) as functions of $q$. The
solid line corresponds to the adiabatic approximation, the line
with squares shows results from simulations. All calculations are
for $m=0, \gamma=1, kT=0.5, F=0.1, D=1$ and $\tau=100/(2 \pi)$.}
\end{figure}

In Fig. 5 we show results from simulations for $J$ and
$\varepsilon$ as functions of $q$ for different values of $D$, the
intensity of the white noise in Eq. (\ref{etapto}). The results
correspond to $T=0$, hence, the only noise present in the system
is the non Gaussian one. On the curve corresponding to the results
for $J$ (Fig. 5.a), we indicate with error bars the dispersion of
the results.

\begin{figure}
\centering
\includegraphics[width=10cm]{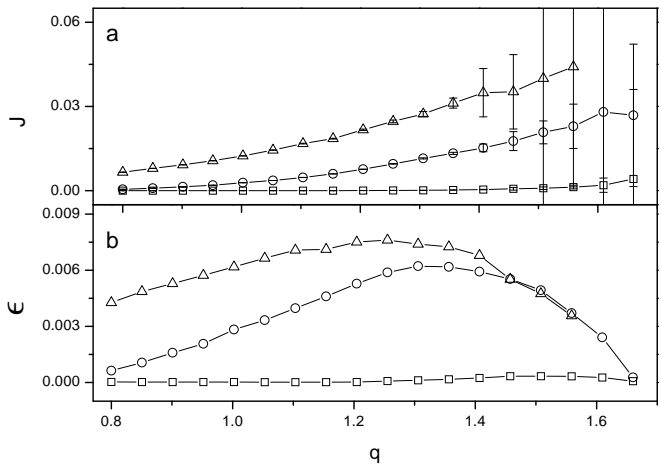}
\caption{Current (a) and efficiency (b) as functions of $q$.
Results from simulations at $T=0$ for $D=1$ (squares), $D=10$
(circles), and $D=20$ (triangles). All calculations are for $m=0,
\gamma=1, F=0.1$ and $\tau=100/(2 \pi)$.}
\end{figure}

It is possible to observe that a huge growth on the fluctuations
of $J$ occurs for the same values of $q$ for which the efficiency
decays. This result induces to think of a connection between the
enhancement of the fluctuations on $J$ and the growth of the width
of $P_q(\eta)$ occurring for $q\to 5/3$ (which, as stated above,
is the origin of the efficiency decay). A simple, however useful,
interpretation of this connection can be given: for $q \to 5/3$,
in spite of having a large (positive) mean value of the current,
for a given realization of the process, the transport of the
particle towards the desired direction is far from being assured
(due to the fluctuations on $J$). Hence, we could say that the
transport mechanism ceases to be efficient.

It is worth recalling here that $q=1$ corresponds to the Gaussian
OU noise \cite{review2}. Hence, these results for constant $D$ and
$\tau$ seem to show that the transport mechanism becomes more
efficient when the stochastic forcing has a non Gaussian
distribution with $q>1$. Let us now discuss the case with inertia
and after that we will return to this aspect through an
alternative approach.

\subsection{Inertial System}

Now we turn to study the $m \neq 0$ case, that is, the situations
in which the inertia effects are relevant. Hence, we consider the
complete form of Eq. (\ref{xpto}). In Fig. 6 we show the
dependence of the current $J$ on the mass $m$ for different values
of $q$ at constant $D$ and $\tau$. The results are from
simulations for zero temperature and without load force. It can be
seen that, as $m$ is increased from $0$, the inertial effects
initially contribute to increase the current, until an optimal
value of $m$ is reached. As expected, for high values of
$m$, the motion of the particle becomes difficult and, for $m\to
\infty$, the current vanishes.

\begin{figure}
\centering
\includegraphics[width=10cm]{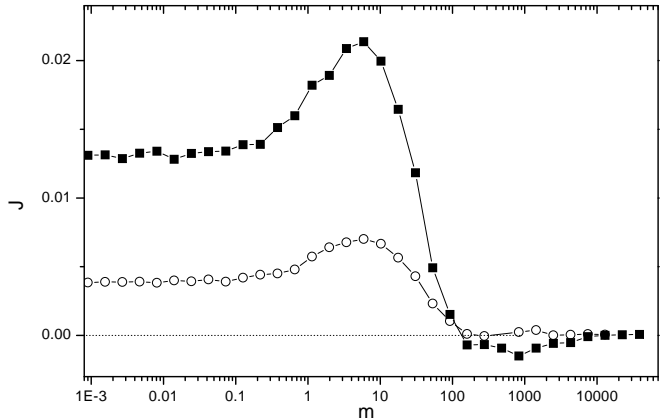}
\caption{Current as function of the mass for $D=1$ and $\tau =
100/2 \pi$ and for different values of $q$: circles $q=1$
(Gaussian noise case), squares $q=1.3$.}
\end{figure}

An interesting effect appears for $q=1.3$. For a well defined
interval of the value of the mass (ranging approximately from
$m=100$ to $m=5000$), a negative current is observed. This can be
explained as a consequence of the high value of the mass, that
makes inertial effects much more important than those of the
ratchet potential. The same velocity's change of sign for high
inertia has been observed in \cite{Lind2} for an OU noise ($q=1$)
for different values of the parameters. However, in the region of
parameters here considered, the effect occurs only if we consider
a non Gaussian noise with a value of $q>1$. In \cite{Lind2}, the
authors give an explanation of this phenomenon in terms of the so
called ``running states'' (i.e. transitions between wells that are
separated by several periods) which is also valid here as was
observed in results from simulations for single realizations of
the process.

These results suggest the possibility that the mass separation
capability of a ratchet system may be in general enhanced by the
inclusion of non Gaussian noises, since we have found mass
separation in a region of parameters where it is not observed for
OU noise. But the separation of masses found here, occurs for
particles with a ratio of masses of the order of $10$ or more (say
$800/80$ in Fig. 6), while it has been shown in \cite{Lind,Lind2}
that, with a load force and considering simply OU noise, particles
of much more similar masses can be separated by ratchets.

In order to compare results, we analyze the same system studied in
\cite{Lind,Lind2} but considering the non Gaussian forcing
described by Eq.(\ref{etapto}). Hence, we study the system in
Eq.(\ref{xpto}) with  $$V(x)=V_2(x)=- \frac{2}{\pi} [\sin(2 \pi
x)+0.25 \sin(4 \pi x)]$$ as the ratchet potential, which is shown
in Fig. 3.b. We focus on the region of parameters where, in
\cite{Lind} (for $q=1$), separation of masses was found. We fix
$\gamma=2, T=0.1, \tau=0.75$, and $D=0.1875$ and assume the
values of the masses $m=m_1=0.5$ and $m=m_2=1.5$ as in
\cite{Lind}. Our main result is that the separation of masses is
enhanced when a non--Gaussian noise with $q>1$ is considered. In
Fig. 7.a we show $J$ as function of $q$ for $m_1=0.5$ and
$m_2=1.5$. It can be seen that there is an optimum value of $q$
that maximizes the difference of currents. This value, which is
close to $q=1.25$, is indicated with a vertical double arrow.
Another double arrow indicates the separation of masses occurring
for $q=1$ (Gaussian OU forcing). We have observed that when the
value of the load force is varied, the difference between the
curves remains approximately constant but both are shifted together
to positive or negative values (depending on the sign of the
variation of the loading). By controlling this parameter it is
possible to achieve, for example, the situation shown in Fig. 7.b,
where, for the value of $q$ at which the difference of currents is
maximal, the heavy ``species'' remains static on average (has
$J=0$), while the light one has $J>0$. Also it is possible to get
the situation shown in Fig. 7.c, at which, for the optimal $q$,
the two species move in opposite directions with equal
average velocity.

\begin{figure}
\centering
\includegraphics[width=12cm]{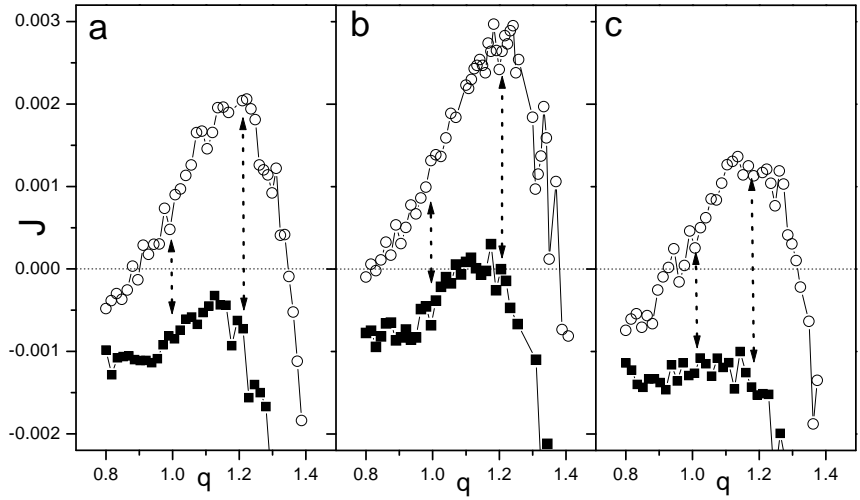}
\caption{Separation of masses: results from simulations for the
current as a function of $q$ for particles of masses $m=0.5$
(hollow circles) and $m=1.5$ (solid squares). Calculations for
three different values of the load force: $F=0.025$ (a), $F=0.02$
(b) and $F=0.03$ (c).}
\end{figure}

\section{Case II: Constant $ D_{ng}$ and $\tau_{ng}$.}

We consider here the second point of view. It corresponds to studying
the behavior of current and efficiency as function of $q$ when
$D_{ng}$ and $\tau_{ng}$ are kept constant (instead of $D$ and
$\tau$ as was done in the previous section). This approach is more
consistent with the image of $\eta$ as a primary (natural) source
of noise, and it isolates the effects of the non-Gaussian
character of the noise distributions by keeping the dispersion and
correlation time at fixed values.

\subsection{Overdamped Case}

We consider Eq. (\ref{xpto}) with $m=0$ and $V(x)$ as in Fig. 3.a
(see Section III.A, overdamped regime). In the simulations, for
each value of $q$, we have adapted the values of $D$ and $\tau$ in
order to obtain the desired values of $D_{ng}$ and $\tau_{ng}$.
This was done by inverting Eq. (\ref{eqn5}) together with the
equation for the best fit for $\tau_{ng}$ (the third one,
presented at the end of Section II).

As $q$ is varied for constant $D_{ng}$, the efficiency is
essentially proportional to the current. Hence, we only present
results for $J$. In Fig. 8 we show the results for the current as
function of $q$ for $\tau_{ng}=2 \pi/100$ and different values of
$D_{ng}$. An interesting result is found for low values of
$D_{ng}$. For $D_{ng}<0.5$ it is observed that the current grows
monotonously with $q$ through most of the range studied (it decays
for $q$ very close to $5/3$) . This means that, for $q>1$ we found
an enhancement of $J$ with respect to the Gaussian noise situation
($q=1$). This enhancement has to be attributed essentially to the
non-Gaussian character of the noise $\eta(t)$, as for every $q$ on
each curve we have considered the same values of dispersion and
correlation time.

\begin{figure}
\centering
\includegraphics[width=10cm]{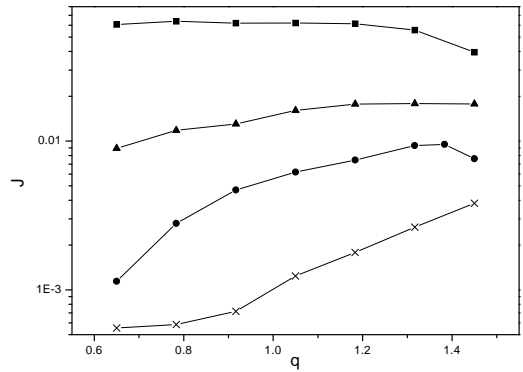}
\caption{Current as a function of $q$ for fixed $\tau_{ng}=100/(2
\pi)$ and different fixed values of $D_{ng}$. From top to bottom,
the curves are for $D_{ng}=1$,$D_{ng}=0.5$,$D_{ng}=0.35$ and
$D_{ng}=0.2$. All calculations are for $m=0, \gamma=1, T=0.1$ and
$F=0.1$.}
\end{figure}

For higher values of $D_{ng}$ the effect disappears: for
($D_{ng}\sim 1$) the dependence of $J$ on $q$ becomes flat in most
of the range analyzed. This means that the non-Gaussian character
does not play a relevant role in this region of parameters, and
the current and efficiency are essentially determined by the
intensity and correlation time of the noise source $\eta$,
independently of the detailed statistical characteristics of the
process $\eta(t)$. For even higher values of $D_{ng} (\sim 2)$ the
optimum value of $q$ that maximizes the current tends toward
values of $q<1$ (results not shown). However in the range of $q$
analyzed the corresponding $J(q)$ curves are essentially flat and
the differences with the $q=1$ case are not distinguishable at all.
Hence, we can remark the enhancement effect occurring for $q>1$ at
low values of $D_{ng}$.

\subsection{Inertial systems}

We consider now the case $m \neq 0$, and study the current as a
function of $q$ for fixed values of $D_{ng}$ and $\tau_{ng}$. We
fix $\gamma=1$ and consider again the potential $V(x)=V_2(x)$
defined in Section III.B.

In Fig. 9 we show the results for $J(q)$ at fixed $D_{ng}
(=0.1875)$ and $\tau_{ng} (=0.75)$, for different values of the
masses and the external force $F$. It can be seen that, when
considering $q \neq 1$, no remarkable enhancement of the mass
separation capabilities of the system is found. Moreover, for
$q>1$ the mass separation effect decreases. However, an
interesting fact is that we find an inversion of current when
considering large enough values of $q$ (depending on the mass and
$F$). This inversion of current is due essentially to the
variation of the non-Gaussian properties of the noise
distribution, as $D_{ng}$ and $\tau_{ng}$ are kept fixed.

\begin{figure}
\centering
\includegraphics[width=9cm]{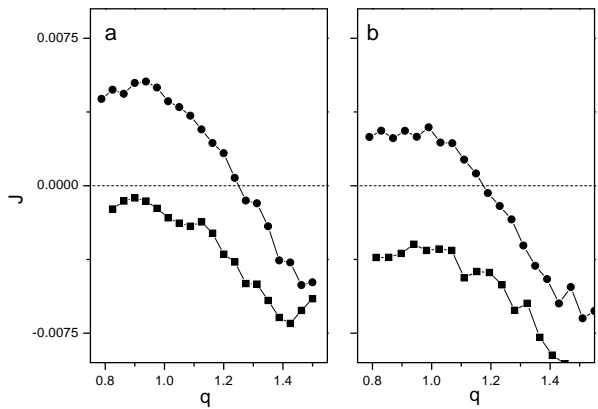}
\caption{Current as a function of $q$ for fixed $\tau_{ng}=0.75$
and $D_{ng}=0.1875$. Figure (a) is for $F=0.015$ and (b) is for
$F=0.025$. In both cases circles correspond to $m=0.5$ while
squares correspond to $m=1.5$.}
\end{figure}

\section{Conclusions}

We have systematically studied the effect of a colored non
Gaussian noise source on the transport properties of a Brownian
motor using two alternative points of view. In the first one, we
analyze the results for the current, efficiency and mass
separation as functions of $q$, for constant values of $D$ and
$\tau$, which are the parameters that could be adequately
controlled, for example, in a designed technological device. The
second point of view corresponds to studying those behaviors as
functions of $q$ when $D_{ng}$ and $\tau_{ng}$ are kept constant.
This is more consistent with the image of $\eta$ as a primary
(natural) source of noise, and it isolates the effects of
the non-Gaussian character of the noise distributions by keeping
the dispersion and correlation time at fixed values.

Considering the first, direct, point of view, what we have found
is that a departure from Gaussian behavior, in particular given by
a value of $q>1$, induces a remarkable increase of the current
together with an enhancement of the motor efficiency. The latter
shows, in addition, an optimum value for a given degree of non
Gaussianity. When inertia is taken into account we have also found
a considerable increment in the mass separation capability.

The second point of view is analogous to the one used in Ref.
\cite{qRE1p} to study stochastic resonance in an
activator-inhibitor system, where it was shown that the
signal-to-noise ratio shows an enhancement as a function of $q$.
Here, by keeping the distribution's width $D_{ng}$ and the
correlation time $\tau_{ng}$ constant, we have compared the
results for ``equivalent" Gaussian and non Gaussian noises. We
have observed that non Gaussian noises with $q>1$ produce an
enhancement of the current when compared to the Gaussian case.
This effect is observed for relatively low values of the noise
intensity $D_{ng}$ and lead us to interpret that, at low values
of $D_{ng}$, the increment of the probability of having arbitrary
high values of the noise that occurs for $q>1$ (with respect to
the Gaussian case) plays a significative role in the determination
of the current. In contrast, for higher values of $D_{ng}$, the
fluctuations dominate the dynamics in such a way that the
Gaussian or non Gaussian character of the noise produces no
relevant differences.

When studying inertial systems at constant $D_{ng}$ and
$\tau_{ng}$ we have not observed relevant effects on the mass
separation capability of the system, in the region of parameters
considered: the effect of mass separation seems to be governed
essentially by the noise intensity and the correlation time.

Another remarkable fact is the occurrence of an inversion of
current as a consequence of varying the parameter $q$ alone
(keeping $D_{ng}$ and $\tau_{ng}$ fixed). This clearly shows the
relevance that the details of the noise distribution may have in
the determination of the transport properties in ratchet systems.
Or, equivalently, how sensitive the ratchet systems could be to
the detailed properties of the noises.

We think that these studies could be of interest for their possible
relation to biologically motivated problems
\cite{inic,nature,caltech} as well as for the potential
technological applications, for instance in ``nanomechanics"
\cite{review1,review2}. More specific studies
on these areas will be the subject of further work. \\ \\

{\bf Acknowledgments:} The authors thank V.Grunfeld for a
revision of the manuscript. Partial support from ANPCyT, Argentine
agency, is acknowledged. HSW wants to thank to the European
Commission for the award of a {\it ``Marie Curie Chair"}, and to
the IFCA and Universidad de Cantabria, Santander, Spain, for the
kind hospitality extended to him.



\begin{thebibliography}{99}

\bibitem{inic} R. D. Vale and F. Oosawa, Adv.\ Biophys.\ {\bf 26},
97 (1990); A. Ajdari and J. Prost, C. R. Acad.\ Sci., Ser.\ II:
Mec., Phys., Chim., Sci.\ Terre Univers.\ {\bf 315}, 1635 (1992).

\bibitem{inic2} R. D. Astumian and M. Bier, Phys.\ Rev.\ Lett.\
{\bf 72}, 1766 (1994);

\bibitem{magnasco} M. Magnasco, Phys. Rev. Lett. {\bf 71} 1477
(1993).

\bibitem{tilting} C. R. Doering, W. Horsthemke, and J. Riordan,
Phys. Rev. Lett. {\bf 72}, 2984 (1994).

\bibitem{tilting2} M.Millonas and M.I Dikman, Phys. Lett. A {\bf
185}, 6 (1994); R. Bartussek, P. H\"anggi and J. G. Kissner,
Europhys.\ Lett.\ {\bf 28}, 459 (1994).

\bibitem{pulsating} P. Reimann and P. H\"anggi, in {\it Lect Notes
in Physics}, vol. 484 (Springer, Berlin 1997).

\bibitem{invers} J. L. Mateos, Phys. Rev. Lett. {\bf 84}, 258
(2000).

\bibitem{review1} R. D. Astumian, Science {\bf 276}, 917 (1997).

\bibitem{review2} P. Reimann, {\it Brownian motors: noisy
transport far from equilibrium}; Phys.\ Rep.\ {\bf 361}, 57
(2002).

\bibitem{qRE1} M.A. Fuentes, R. Toral and H.S. Wio, Physica A {\bf
295}, 114-122 (2001); M.A. Fuentes, H.S. Wio and R. Toral, Physica
A {\bf 303}, 91--104 (2002).

\bibitem{qRE1p} M.A. Fuentes, C. Tessone, H.S. Wio and R. Toral,
Fluctuations and Noise Letters {\bf 3}, L365 (2003).

\bibitem{qRE2} F.J. Castro, M.N. Kuperman, M.A. Fuentes and H.S.
Wio, Phys. Rev. E {\bf 64}, 051105 (2001).

\bibitem{qruido1}  H.S. Wio, J.A. Revelli and A.D. S\'anchez,
Physica D {\bf 168-169}, 165-170 (2002).

\bibitem{qruido2} H.S. Wio and R. Toral, in {\it Anomalous
Distributions, Nonlinear Dynamics and Nonextensivity}, H. Swineey
and C. Tsallis (Eds.), Physica D (in press).

\bibitem{qruido3} H.S. Wio, {\it On the Role of Non-Gaussian
Noises}, chapter in Ref. (\cite{tsallis2}).

\bibitem{tsallis} C. Tsallis, Stat. Phys. {\bf 52}, 479 (1988);
E.M.F. Curado and C. Tsallis, J. Phys. A {\bf 24}, L69 (1991);
ibidem, {\bf 24}, 3187 (1991); ibidem, {\bf 25}, 1019 (1992).

\bibitem{tsallis2} M.Gell-Mann and C. Tsallis (Eds.), {\it
Nonextensive Entropy-Interdisciplinary Applications} (Oxford U.P.,
Oxford, 2003).

\bibitem{nature} S. M. Bezrukov and I. Vodyanoy, Nature {\bf 378},
362 (1995); D. Nozaki, D.J. Mar, P. Griegg and J.D. Collins, Phys.
Rev. Lett. {\bf 72}, 2125 (1999).

\bibitem{caltech} A. Manwani and C. Koch, Neural Comp.,  {\bf 11},
1797 (1999); A. Manwani, PhD Thesis, CALTECH, (2000).

\bibitem{Seki} K. Sekimoto, Prog. Theor. Phys. Supl. {\bf 130}
17 (1998); J.M.R. Parrondo, et al., Europhys. Lett. {\bf 43}
248 (1998).

\bibitem{comFP} In doing this, we use a piecewice linear
approximation for the ratchet potential which is
$V(x)=2.5-(5/1.1)x$ for $x<1.1$ and $V(x)=-2.5+[5/(2 \pi
-1.1)](x-1.1)$ for $1.1<x<2\pi$. We solve the Fokker--Planck
equation for $0\le x \le 2 \pi$ with periodic boundary condition
asking for continuity of the distribution and the current at
$x=1.1$.

\bibitem{new} T.E. Dialynas, K. Lindenberg and G.P. Tsironis,
Phys. Rev. E {\bf 56}, 3976 (1997).

\bibitem{Lind} R. Lindner, L. Schimansky-Geier, P. Reimann, and P.
H\"anggi, in {\it Applied Nonlinear Dynamics and Stochastic
Systems Near the Millenium} pg. 309, J.B. Kadtke and A. Bulsara,
Eds. (AIP, 1997).

\bibitem{Lind2} R. Lindner, L. Schimansky-Geier, P. Reimann, P.
H\"anggi and M. Nagaoka, Phys. Rev. E {\bf 59}, 1417 (1999).

\bibitem{Landa} P.S Landa, Phys. Rev. E {\bf 58}, 1325 (1998).

\end{thebibliography}
\end{document}